\documentclass[jkps,twoside,twocolumn,fleqn,showpacs,showkeys,showdoi]{revtex4}
\usepackage[pdftex]{graphicx}
\usepackage{amssymb}
\usepackage{amsmath}
\usepackage{bm,multirow}
\usepackage{eurosym}
\usepackage{enumerate}
\usepackage{slashbox}
\usepackage{subeqnarray}

\newcommand{\nn}{\nonumber}

\def\be{\begin{equation}}
\def\ee{\end{equation}}
\def\ba{\begin{eqnarray}}
\def\ea{\end{eqnarray}}
\volumenumber{68} \issuenumber{6} \volumeyear{March 2016}
\startpage{0}
\endpage{0}

\newcommand{\Rmnum}[1]{\expandafter\@slowromancap\romannumeral #1@}

\begin{document}
\setcounter{page}{1}

\title[]{Quantum Corrections to the Hawking Radiation Spectrum}

\author{Youngsub \surname{Yoon}}
\email{youngsuby@snu.ac.kr}
\affiliation{Department of Physics and
Astronomy, Seoul National University, Seoul
        08826, Korea}

\date[]{Received 22 January 2016, in final form 19 February 2016}

\begin{abstract}
In 1995, Bekenstein and Mukhanov suggested that the Hawking
radiation spectrum was discrete if the area spectrum was quantized
in such a way that the allowed areas were integer multiples of a
single unit area. However, in 1996, Barreira, Carfora, and Rovelli
argued that the Hawking radiation spectrum was continuous if the
area spectrum was quantized with an infinite number of unit areas,
as predicted by loop quantum gravity, rather than quantized with the
single unit area considered by Bekenstein and Mukhanov. In this
paper, contrary to what Barreira, Carfora, and Rovelli argued, we
show that the Hawking radiation spectrum is still discrete when the
area spectrum is quantized as loop quantum gravity predicts. In
particular, we show that, for a black hole of a given temperature,
the Hawking radiation spectrum is truncated at frequencies below a
certain frequency.
\end{abstract}



\maketitle

\section{Introduction}

Thanks to Hawking, a black hole is well known to emit
particles\cite{Hawking}. Hawking also argued that the radiation
spectrum must follow the Planck radiation spectrum. However,
strictly speaking, this is not necessarily the case as Hawking's
calculation was semi-classical rather than fully quantum. Perhaps
starting from this observation, Bekenstein and Mukhanov showed that
the Hawking radiation spectrum is discrete if the allowed area
values are integer multiples of a single unit area\cite{Bekenstein}.
In this paper, by closely reviewing how Planck's black body
radiation formula is derived, we argue that the Hawking radiation
spectrum is discrete not just when the allowed area values are
integer multiples of a single unit area but also when the area
spectrum is quantized as loop quantum gravity
predicts\cite{Discreteness, General, Ashtekar}. This result
contradicts previous arguments by Barreira \textit{et al.}\cite{BCR}
and Krasnov\cite{Krasnov}.

In Section II, we closely review the arguments that the Hawking
radiation spectrum is continuous. Then, we propose a selection rule
for quantum black holes, namely, that the area of the black hole can
decrease only by an amount equal to a unit area upon emission of a
photon. In Section III, by relating the area reduction with the
energy of the emitted photon, we show that the selection rule
requires that the Hawking radiation spectrum be discrete. Also in
that section, we show that the Hawking radiation spectrum is
truncated below a certain photon energy. We consider three different
area spectra: the isolated horizon framework\cite{isolated, Ghosh},
the area spectrum of Tanaka and Tamaki\cite{Tanaka}, and the area
spectrum of Kong and Yoon\cite{blackholeentropy, ToAppear}. For all
cases, we find that Hawking radiation is discrete. In section IV, we
present an alternative, but equivalent, derivation of the relation
between the area reduction and the energy of the emitted photon,
which again leads to discreteness. In Section V, we derive the
selection rule proposed in Section II. In Section VI, we study how
our result changes if we consider logarithmic corrections to the
Bekenstein-Hawking entropy. In Section VII, we compare our results
with those of Refs. 13 and 14. In particular, we show that our
results are clearly different. In Section VIII, we conclude our
article.

\section{Selection rules for quantum black holes}

In their article\cite{BCR} Barreira \textit{et al.} notice that
``the spacing of the energy levels [of a black hole] decreases
\emph{exponentially} with M''. They go on to say, ``It follows that
for a macroscopical black hole the spacing between energy levels is
infinitesimal, and thus the spectral lines are virtually dense in
frequency''. Their argument that for a macroscopic black hole the
spacing between energy levels is infinitesimal is correct.
Nevertheless, their argument that the spectral lines are virtually
dense in frequency is wrong. They assume that a photon emitted from
a black hole can have any energy as long as that energy can be
written as the difference between the two energy values an arbitrary
black hole can have. They implicitly assume that a black hole can
turn into any other black hole with less energy. Similarly, they
implicitly assume that a black hole can turn into any other black
hole with less area.

Let's phrase their argument mathematically. Let's say that we have
the following area eigenvalues (\textit{i.e.}, the unit areas):
\begin{equation}
A_i=A_1, A_2, A_3, A_4, A_5, A_6....
\end{equation}
Then, the black hole area $A$ must be given by the following
formula:
\begin{equation}
A=\sum_i N^i A_i\label{NjAj},
\end{equation}
where the $N^i$s are non-negative integers. Here, we can regard the
black hole as having $\sum N^i$ partitions, each of which has one of
the $A_i$ as its area. In this mathematical language, we can express
the consideration of Barreira \textit{et al.} as follows: the black
hole with initial area $A_{\mathrm{int}}=\sum N^i_{\mathrm{int}}
A_i$ can turn into a black hole with final area
$A_{\mathrm{fin}}=\sum N^i_{\mathrm{fin}} A_i$ through the emission
of photons, as long as $A_{\mathrm{fin}} < A_{\mathrm{int}}$,
without any restrictions on the set of $N^i_{\mathrm{fin}}$.

However, if we assume that the emission of a photon is local, this
is not the case. For a photon to be emitted locally, it should be
emitted from a single area quantum, not simultaneously from multiple
area quanta separated in space. Possibly following these
considerations, Krasnov argued that\cite{Krasnov}
\begin{quote}
``Consider a quantum process in which the black hole jumps from a
state $|\Gamma \rangle$ to state $|\Gamma' \rangle$, such that the
horizon area changes. This, for example, can be a process in which
one of the flux lines piercing the horizon breaks, with one of the
ends falling into the black hole and the other escaping to infinity
(see Fig. 1b). This is an example of the emission process; the two
ends of the flux line can be thought of as the two particle
anti-particle quanta in Hawking's original picture [6] of the black
hole evaporation.''
\end{quote}
Translating this into a mathematical formula, what Krasnov argues is
the following:
\begin{equation}
\Delta A= A_j-A_i\label{Krasnov}
\end{equation}
for some $A_i > A_j$. In other words, the partition with area $A_i$
on the black hole horizon shrinks into a partition with area $A_j$
upon the emission of a particle because the anti-particle reaches
this partition of the black hole horizon.

However, Krasnov's argument is also troublesome. In Section V, we
will explain why the selection rule should be
\begin{equation}
\Delta A=-A_i\label{ourformula}
\end{equation}
for some $i$. Before doing so, we will explain the consequences of
Eq. (\ref{ourformula}) in the next two sections.

\section{The discreteness of the Hawking radiation spectrum}

From black hole thermodynamics, we know the following\cite{Hawking}:
\begin{equation}
r=2M,
\end{equation}
\begin{equation}
A=4\pi r^2=16\pi M^2,\label{AM}
\end{equation}
\begin{equation}
kT=\frac{1}{8\pi M},\label{kT}
\end{equation}
where $A$ is the horizon area of the black hole, $T$ its
temperature, $r$ its radius, and $M$ its mass, and $k$ is
Boltzmann's constant. Here, we consider the case of a Schwarzschild
black hole for simplicity, but it can easily be generalized to the
generic case as is done in Section IV.

Now consider the emission of a photon from the black hole. As the
photon is emitted, the black hole loses energy; thus, its area
decreases by $A_{i}$, the unit area predicted by loop quantum
gravity as we argued in Eq. (\ref{ourformula}) in the last section.
From this consideration, we can calculate $E_{photon}$, the energy
of the emitted photon. First of all, the mass of the black hole
decreases as
\begin{equation}
\Delta M=-E_{photon}
\end{equation}
Then, considering Eqs. (\ref{AM}) and (\ref{kT}), the area of the
black hole decreases as
\begin{eqnarray}
\Delta A&=& 32 \pi M \Delta M=-32 \pi M E_{photon}\nn \\ &=&-\frac{4
E_{photon}}{k T}=-A_{i},
\end{eqnarray}
where in the last step, we assert that the black hole area must be
decreased by the unit area $A_{i}$ predicted by loop quantum
gravity. Therefore, we conclude the following:
\begin{equation}
E_{photon}=\frac{A_{i}}{4} k T. \label{recover}
\end{equation}
Here, we see easily that the energy of the emitted photon is
quantized because $A_{i}$ is quantized. In particular, as loop
quantum gravity predicts that a non-zero minimum area exists, a
non-zero energy exists for the photons emitted from a black hole of
a given temperature.

In the case of the isolated horizon framework\cite{isolated, Ghosh},
the minimum area is given by $4\pi\sqrt{3} \gamma$ where $\gamma$ is
the Immirzi parameter. Therefore, we have the following for the
minimum energy of the emitted photon:
\begin{equation}
E_{min}\approx 1.49 k T
\end{equation}
(see Fig.~1). The Hawking radiation is truncated below this energy.
The discrete frequency values allowed for Hawking radiation are
represented by solid lines. In the case of the Tanaka-Tamaki
scenario\cite{Tanaka}, the minimum area is given by $4 \pi \gamma$,
where $\gamma$ is the Immirzi parameter for this case. This gives
the following for the minimum energy of emitted photon:

\begin{equation}
E_{min}\approx 2.462 k T
\end{equation}
(see Fig.~2). In the case of the Kong-Yoon
scenario\cite{blackholeentropy, ToAppear}, the minimum area is given
by $4 \pi \sqrt{2}$. Therefore, we have the following minimum
energy:
\begin{equation}
E_{min}\approx 4.44 k T
\end{equation}
(see Fig.~3).

\begin{figure}[t!]
\includegraphics[width=7.7cm]{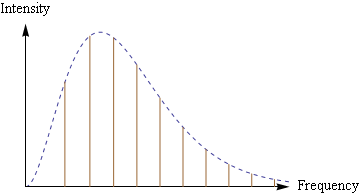}
\caption[0]{(Color online) Isolated horizon framework.}
\end{figure}


\begin{figure}[t!]
\includegraphics[width=7.7cm]{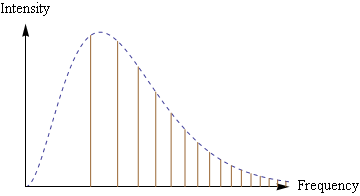}
\caption[0]{(Color online) Tanaka-Tamaki scenario.} \label{Fig2}
\end{figure}



\begin{figure}[t!]
\includegraphics[width=7.7cm]{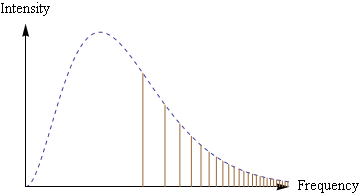}
\caption[0]{(Color online) Kong-Yoon scenario.} \label{Fig3}
\end{figure}

\newpage

\section{Alternative derivation}

In this section, we present a simpler derivation. From
thermodynamics, we have the following:

\begin{equation}
\Delta Q=T \Delta S\label{dQTdS}
\end{equation}
Plugging in the equalities
\begin{eqnarray}
&& \Delta Q= -E_{photon},\label{dQ}\\
&& \Delta S= -k A_{i}/4,
\end{eqnarray}
we recover Eq. (\ref{recover}).

\section{Griffiths' quantum mechanics and Pathria's statistical mechanics}
In his famous textbook\cite{Griffiths}, Griffiths considers a
statistical mechanics problem as follows:
\begin{quote}
``Now consider an arbitrary potential, for which the one-particle
energies are $E_1, E_2, E_3,\cdots,$ with degeneracies
$d_1,d_2,d_3,\cdots$. Suppose we put $N$ particles into this
potential; we are interested in the configuration
($N_1,N_2,N_3,\cdots$), for which there are $N_1$ particles with
energy $E_1$, $N_2$ particles with energy $E_2$ and so on. How many
different ways can this be achieved?''
\end{quote}
Then, he shows that the answer is given by the following for the
case of bosons:
\begin{equation}
Q=\prod_{n=1}^{\infty}\frac{(N_n+d_n-1)!}{N_n!(d_n-1)!}.
\end{equation}
He also explains that we have the following two conditions:
\begin{equation}
\sum_{n=1}^{\infty}N_n=N,\qquad\sum_{n=1}^{\infty}N_nE_n=E.\label{twoconditions}
\end{equation}
The first condition requires that the total number of particles is
$N$ while the second requires that the total energy is $E$. To find
the most probable configuration $(N_1,N_2,N_3,\cdots)$, he maximizes
$\ln Q$ as follows:
\begin{equation}
G\equiv \ln
Q+\alpha\left[N-\sum_{n=1}^{\infty}N_n\right]+\beta\left[E-\sum_{n=1}^{\infty}N_nE_n\right],\label{G}
\end{equation}
where $G$ is to be maximized and $\alpha$ and $\beta$ are Lagrange
multipliers. He concludes that
\begin{equation}
N_n=\frac{d_n}{e^{\alpha+\beta E_n}-1}.\label{Nn}
\end{equation}
Of course, in the case of photons, the number $N$ is not conserved,
so we set $\alpha=0$ in Eqs. (\ref{G}) and (\ref{Nn}). Furthermore,
we know $\beta=1/(kT)$, which implies
\begin{equation}
N_n=\frac{d_n}{e^{E_n/(kT)}-1}.\label{Nnplugged}
\end{equation}

We also know (see, \textit{e.g.}, Section 6.4 of ``Statistical
Mechanics'' by Pathria) that the intensity $I(E_n)$ of photons
emitted through black body radiation is given by

\begin{eqnarray}
I(E_n) &=& \frac c4 N_n A = \frac c4 \frac{d_n}{e^{E_n/(kT)}-1} A\nn
\\ &=& \frac c4\frac{8\pi f^2 df}{e^{hf/(kT)}-1} A.
\end{eqnarray}
In the last two expressions, we have substituted the density of
states for the degeneracy $d_n$ in the numerator and written the
photon energy $E_n$ in terms of the frequency $f$ as $h f$.
Recalling that the black hole (or any black body) loses energy $hf$
upon emission of a photon with frequency $f$, we can write
\begin{equation}
\Delta E=-hf;
\end{equation}
thus,
\begin{equation}
\Delta E=-E_n.\label{DeltaEEn}
\end{equation}
This equation shows that only the radiation associated with $E_n$
(\textit{i.e.}, a single area quanta or $n$th unit area) is
possible. This can be seen better by noticing that the second
equation of Eq. (\ref{twoconditions}) runs parallel with Eq.
(\ref{NjAj}). They are actually related by Eq. (\ref{recover}).
Therefore, we derived Eq. (\ref{ourformula}) (\textit{i.e.}, $\Delta
A=-A_i$).

Now, suppose a hypothetical case in which the area deduction is
given by $\Delta A=A_j-A_i$ as Krasnov argued. In such a case, we
would have $\Delta E=E_j-E_i$, which implies that the energy of the
emitted photon is given by $hf=E_i-E_j$. Given this, let's compare
the black body radiation formula in this hypothetical case with Eq.
(\ref{Nnplugged}). The denominator does not match as Eq.
(\ref{Nnplugged})'s denominator is $e^{E_n/(kT)}-1$ while Krasnov's
hypothetical one would be $e^{(E_i-E_j)/(kT)}-1$. They are clearly
different. Furthermore, the numerator does not match either. In the
case of Eq. (\ref{Nnplugged}), we have the degeneracy of the $n$th
quanta given as $d_n$. In Krasnov's hypothetical case, whether the
degeneracy should be $d_i$ or $d_j$ or $d_id_j$ is not clear.
Perhaps no consistent way exists to assign a value to the numerator
such that it reduces to $d_n$ in the case where $E_i=E_n$ and
$E_j=0$ but is different from $d_n$ when $E_i=E_n$ but $E_j\neq0$.
In conclusion, Krasnov's area deduction condition is wrong as it
cannot reproduce Eq. (\ref{Nnplugged}).

\newpage
\section{logarithmic corrections to the black hole entropy}

In the presence of logarithmic corrections to the black hole
entropy, Eq. (\ref{recover}), is modified. In this section, we
consider the fully $SU(2)$ framework as an example. Other cases can
be dealt with in a similar way. In the fully $SU(2)$ framework, we
have\cite{LivineTerno,ENP1, ENP2, ENP3}:
\begin{equation}
S=\frac{A}{4}-\frac{3}{2}\ln A+O(1).
\end{equation}
Given this and using Eqs. (\ref{AM}), (\ref{kT}), (\ref{dQTdS}), and
(\ref{dQ}), we obtain
\begin{equation}
E_{photon}=\left(\frac{ k T}{4}-6\pi (kT)^3\right) A_{i}.
\end{equation}

\section{Comparison with the D\'{\i}az-Polo-Fern\'{a}ndez-Borja effect}

In Refs.~13 and 14, an argument is made that loop quantum gravity
effects modify the Hawking radiation spectrum, though differently
from the results presented in this paper. In this section, we
present their reasoning and clearly show that our results are
different from theirs.

Considering the isolated horizon framework, the authors of Ref. 14
note that $\Delta A$ in Eq. (\ref{Krasnov}) in our paper tends to
contain integer multiples of $2.41\cdots$ more often than other
values, though other values occur as well. Therefore, the
corresponding photon energy associated with the values of $\Delta A$
is peaked in the spectrum around integer multiples of $2.41\cdots$
while photon energies from other values of $\Delta A$ form a
continuous background. Again, from the following sentence, what they
clearly considered in their paper is not our formula in Eq.
(\ref{ourformula}), but Krasnov's formula (Eq. (\ref{Krasnov}) in
our paper):
\begin{quote}
``In our analysis we are assuming that a black hole can undergo a
transition from any configuration to any other one with the only
condition that the final state belongs to a lower area band.''
\end{quote}
They conclude their paper as follows:
\begin{quote}
``...the physical consequence that one can extract when studying the
spectroscopy is not the discretization of the radiation spectrum and
the appearance of a minimum emission frequency (as in the
Bekenstein-Mukhanov scenario). The imprint of quantum gravity
effects in Hawking radiation (for microscopic black holes) within
this framework is manifested in the emergence of some equidistant
brighter lines over a continuous background spectrum.''
\end{quote}
We see here that their results are different from ours. In our case,
we have the discretization of the radiation spectrum and the
appearance of a minimum emission frequency, even though we used
multiple unit areas as predicted by loop quantum gravity (unlike the
Bekenstein-Mukhanov scenario). In our case, the discrete lines are
not equidistant because the $A_{i}$ values are not equidistant nor
do we have a continuous background spectrum. In conclusion, we want
to note that the results refs. 13 and 14 seem to be wrong, as they
are based on Eq. (\ref{Krasnov}), Krasnov's incorrect formula.

\section{Discussion and Conclusions}

In this article, by closely following an elementary result explained
in a quantum mechanics textbook, we showed that the Hawking
radiation spectrum must be discrete and must be truncated below a
certain frequency, deviating from the simple Planck radiation
spectrum. We also want to note that Brian Kong and the author have
given strong evidence for this phenomenon in two other
papers\cite{blackholeentropy, ToAppear}; we calculated a new area
spectrum based on what we called ``newer'' variables and calculated
the discrete Hawking radiation spectrum. Then, we approximated the
new Hawking radiation spectrum as being continuous, but truncated,
below a certain frequency predicted by the ``newer'' variables. This
approximation is reasonable if you look at Fig. 3. Using this
approximation, we obtained $172.87\cdots$ for a certain value
associated with the strength of the Hawking radiation spectrum. We
also estimated this value to be 172$\sim$173 by using a
\emph{totally different} method that involved statistical fitting.
Again, this should be regarded as strong evidence for the
discreteness of the Hawking radiation spectrum, as one would not
have had this numerical agreement if there had not been a minimum
frequency for the Hawking radiation spectrum. In any case, we hope
that the discreteness of the Hawking radiation spectrum will be
uncontroversially confirmed by detecting and measuring Hawking
radiation at the Large Hadron Collider.

\section{Addendum: implications on the black hole information paradox}
Notice that Eq. (\ref{recover}) implies that the Hawking radiation is purely thermal. Therefore, the information on the objects that have fallen into a black hole cannot be retrieved through the Hawking radiation.

~\\

~\\

~\\

\begin{acknowledgments}

We thank Carlo Rovelli for helpful discussions, especially for the
idea of locality. This work was supported by National Research
Foundation of Korea (NRF) grants 2012R1A1B3001085 and
2012R1A2A2A02046739.

\end{acknowledgments}

~\\

~\\~\\

\end{document}